%% file: root.tex
\documentclass[letterpaper, 10 pt, conference]{ieeeconf} 
\usepackage{tabularx}
\usepackage{amsmath}
\usepackage{amssymb}
\usepackage{graphicx}
\usepackage{tikz}
\usepackage{svg}
\usepackage{pdfpages}
\usepackage{subcaption}
\usepackage{dblfloatfix} 

\IEEEoverridecommandlockouts  

\author{Ye Tao$^1$, Hongyi Wu$^2$, Ehsan Javanmardi$^1$, Manabu Tsukada$^1$, Hiroshi Esaki$^1$
\thanks{*These research results were obtained from the commissioned research Grant number 01101 by the National Institute of Information and Communications Technology (NICT), Japan.}
\thanks{*This work was partly supported by JSPS KAKENHI (grant numbers: 21H03423).}
\thanks{$^{1}$ Graduate School of Information Science and Technology, The University of Tokyo. %
{\newline\tt\small tydus@hongo.wide.ad.jp}, %
} %
\thanks{$^{2}$ School of Mathematics and Physics, University of Queensland. {\tt \small hongyi.wu1@uq.net.au}}%
}

\usetikzlibrary{shapes,arrows}
\usetikzlibrary{shapes.geometric,shapes.misc,arrows.meta,calc,fit,positioning}

\usepackage{hyperref}

\title{\LARGE \bf Zero-Knowledge Proof of Distinct Identity: a Standard-compatible Sybil-resistant Pseudonym Extension for C-ITS}

\begin{document}

\maketitle

\begin{abstract}
Pseudonyms are widely used in Cooperative Intelligent Transport Systems (C-ITS) to protect the location privacy of vehicles.
However, the unlinkability nature of pseudonyms also enables Sybil attacks, where a malicious vehicle can pretend to be multiple vehicles at the same time.
In this paper, we propose a novel protocol called zero-knowledge Proof of Distinct Identity (zk-PoDI,) which allows a vehicle to prove that it is not the owner of another pseudonym in the local area, without revealing its actual identity.
Zk-PoDI is based on the Diophantine equation and zk-SNARK, and does not rely on any specific pseudonym design or infrastructure assistance.
We show that zk-PoDI satisfies all the requirements for a practical Sybil-resistance pseudonym system, and it has low latency, adjustable difficulty, moderate computation overhead, and negligible communication cost.
We also discuss the future work of implementing and evaluating zk-PoDI in a realistic city-scale simulation environment.

\end{abstract}

\section{Introduction}
\label{sec:introduction}
Road transportation has been one of the most essential services for human mobility since ancient times.
Intelligent Transportation Systems (ITS) aim to improve road transportation efficiency, safety, and sustainability.
Cooperative ITS (C-ITS) enables real-time information sharing between vehicles and surrounding objects like roadside units and pedestrians through vehicle-to-everything (V2X) communication.
International organizations like ISO and ETSI define standards for interoperability of V2X communication in C-ITS. These standards are crucial for harmonized development and deployment of C-ITS technologies.

Pseudonyms, or temporary identifiers, are essential for protecting privacy and security in C-ITS communications.
They allow vehicles to exchange information without revealing true identities, maintaining user anonymity and preventing unauthorized tracking or access.
In C-ITS, pseudonyms are often government-issued short-term digital certificates, ensuring legal identities while protecting vehicle privacy.
ISO~\cite{Iso2022-zc} and ETSI~\cite{Etsi2022-xf} standards mandate that pseudonym designs meet unlinkability requirements~\cite{Pfitzmann2010-xo}, meaning no party except law enforcement agencies (LEAs) and owners can link two pseudonyms to the same vehicle.

However, the legality and unlinkability of pseudonyms make the system vulnerable to Sybil attacks~\cite{Douceur2002-vf}.
A Sybil attack involves a malicious entity transmitting data using multiple pseudonyms simultaneously, pretending to be different vehicles.
This can disseminate false information, disrupt traffic flow, or engage in malicious activities, threatening transportation safety and functionality.
Works relying on voting or cross-verification, like zk-PoT~\cite{Tao2023-xq}, are especially vulnerable.
Multiple sybil-resist pseudonyms are proposed~\cite{Maram2021-bp,Friebe2016-if,Florian2015-po,Li2021-rk,Akil2023-ka} but require global databases, blockchains, or prior knowledge like social graphs, which are impractical for C-ITS due to its locality and intermittent connectivity.

For vehicular ad-hoc networks (VANETs), vehicles cannot always have global or infrastructure connections.
Preloading pseudonyms on vehicles becomes a viable approach, but creates difficulties for law enforcement agencies (LEAs) in managing pseudonym lifecycles, resulting in multiple valid pseudonyms simultaneously.

In C-ITS, every pseudonym is a valid ID, vehicles can have multiple pseudonyms simultaneously, and pseudonym owners are indistinguishable publicly.
These features facilitate Sybil attacks by enabling simultaneous pseudonym usage.
However, if the public could determine the owner of pseudonyms, it would break the unlinkability.
This leaves us with a paradox: how can the vehicle give out just the right amount of knowledge, to convince the others, but not compromise the unlinkability?

We observed that in C-ITS, vehicles don't need outdated or distant information.
They don't care if a nearby vehicle also appears far away.
Among conflicting C-ITS demands, compromising global Sybil-resistance has the least impact.
Therefore, we abandoned pursuing global Sybil-resistance and instead addressed local Sybil-resistance.
Vehicles no longer need to prove uniqueness from every vehicle, but only the distinctiveness from surrounding vehicles.
We call this ``\textbf{the locality assumption}.''

Based on the locality assumption, we proposed a novel method called zero-knowledge Proof of Distinct Identity (zk-PoDI), an extension to the pseudonym systems.
zk-PoDI enables a vehicle to prove that it is \textit{not} the owner of a given pseudonym, while not breaking the unlinkability property of the pseudonyms.
Utilizing the zk-PoDI, a vehicle can prove it is not another vehicle whenever needed, i.e. it is not potentially conducting a Sybil attack, thus its identity is legit and its shared information is trustworthy.
Zk-PoDI has very few assumptions underlying pseudonyms and thus can be applied to a vast amount of pseudonym systems as an extension.

Our contributions are as follows:
\begin{enumerate}
    \item Based on the observation of the locality, we relaxed the requirement of global Sybil-resist of the pseudonym and thus came out with an approach of proving locally distinctive.
    \item Exploiting the Diophantine equation, we designed an open-problem-backed construction to prove the distinctiveness of a vehicle, the distinctive identity criterion (DIC).
    \item By utilizing the zero-knowledge succinct non-interactive argument of knowledge (zk-SNARK), we convert the DIC into a zero-knowledge prove mechanism, to protect the user's pseudonymity by hiding the actual secret orthonym.
    \item Based on the above mathematical model, we proposed the actual protocol zk-PoDI, which can be integrated into most pseudonym designs as an extension.
\end{enumerate}

The rest of the paper is organized as follows.
Section~\ref{sec:related} introduces the related work.
Section~\ref{sec:problem} shows the problem statement, makes some assumptions about the pseudonym, and raises some requirements.
Section~\ref{sec:math} recalls several mathematical tools.
Section~\ref{sec:proposal} starts with the mathematical solution to the problem, then proposes the actual protocol, zk-PoDI, and shows how can it prevent the Sybil attack while not compromising privacy.
Section~\ref{sec:discussion} discusses the performance of zk-PoDI and its resistance against several attacks.
Finally, section~\ref{sec:conclusion} concludes the paper.

\section{Related Work}
\label{sec:related}
\subsection{Pseudonyms in ITS}
In the literature, many works are proposing a new type of pseudonym to better adapt to the application scenarios of C-ITS.
They also keep security in mind and show strength against various attacks.
Mundhe et al.~\cite{Mundhe2020-cx} proposes an efficient Lightweight Ring Signature-Based Conditional Privacy-Preserving Authentication (RCPPA) scheme for VANETs.
It employs pseudonym-based authentication and lattice-based ring signatures to ensure message authentication without revealing the real identity of vehicles.
%
Ali and Li~\cite{Ali2020-fx} introduce an efficient Identity-based Conditional Privacy-Preserving Authentication (ID-CPPA) signature scheme for V2I communication in VANETs.
Addressing the inefficiency in current schemes, the proposed ID-CPPA scheme utilizes bilinear maps and general one-way hash functions, supporting batch signature verification at RSUs.

From the above examples, we can see that the current proposed works about pseudonyms do not directly have a built-in mechanism against the Sybil attacks.
It is thus crucial to propose some mechanism for Sybil resistance or Sybil detection to extend those works.

\subsection{Sybil attack and its countermeasures}
C-ITS is especially vulnerable to Sybil attack because of its pseudonym design.
Many works, either Sybil-resist pseudonyms or the mechanisms helping other pseudonyms block Sybil attacks, were proposed.
Baza et al.~\cite{Baza2022-fe} propose a C-ITS Sybil-attack detection scheme that utilizes proofs of work and location, based on RSU-issued timestamps, featuring a high detection rate and low false negatives and acceptable communication and computation overhead.
Trauernicht et al.~\cite{Trauernicht_undated-bs} introduce a standard aligned concept for the long-term exclusion of Sybil attackers in C-ITS while preserving the privacy of benign stations.
The proposed mechanism prevents the reuse of Authorization Tickets, ensures no false exclusions of benign stations, and maintains privacy during misbehavior evaluation.

 

\subsection{Zero-knowledge proofs}
Zero-knowledge proofs were introduced by S. Goldwasser, S. Micali, and C. Rackoff in 1989~\cite{Goldwasser1989-js}. They also introduced interactive proof systems in the same article. The primary objective of a zero-knowledge proof is for the prover to demonstrate the truth of a statement (e.g., possessing the solution to an equation) to the verifier without disclosing any information beyond the statement's validity. Zero-knowledge proofs must satisfy the following three conditions:

\textbf{Completeness}: If the statement is true, an honest verifier should be convinced of the prover's claim.

\textbf{Soundness}: If the statement is false, there should be only a small probability that the verifier erroneously accepts the dishonest prover's claim.

\textbf{Zero Knowledge}: Throughout the process, the verifier should learn nothing except the truth of the statement.

Zero-knowledge proofs have a wide range of applications in many fields, such as in ML-as-a-service (MLaaS)~\cite{kang2022scaling}. In their work, the authors addressing the challenge of ensuring correct predictions in the presence of potentially malicious or faulty service providers by using Zero-knowledge proofs. 

Recently, there has been a growing focus on utilizing zero-knowledge proofs to enhance the security and anonymity of C-ITS. In~\cite{9809980}, the authors introduced a novel and efficient anonymous authentication approach for the Internet of Vehicles (IoV). 
In the realm of wearable-based Collaborative Indoor Positioning Systems, privacy concerns prompted the proposal of a decentralized Attribute-based Authentication (ABA) protocol in~\cite{CASANOVAMARQUES2023100801}. 

\section{Problem Statement, Assumptions, Requirements and Approach}
\label{sec:problem}
In the last section, we reviewed the current progress of the pseudonym systems and Sybil attack countermeasures in C-ITS, with their reasons against deployment respectively.
In order to fill the gap in pseudonyms against the Sybil attack, in this section, we will formalize our particular problem regarding the Sybil attack, make a few reasonable assumptions about the underlying pseudonym system, and finally detail our requirements for the proposed work.

\subsection{Problem statement}


To enhance the efficiency of road transportation, we aim for vehicles within a local area to exchange information and report their respective situations.
Each vehicle should utilize multiple pseudonyms with no obvious connections to protect the owner's privacy.

Consider a vehicle, Alice.
It will receive messages from different pseudonyms. 
Alice can adjust its driving strategy based on these messages.
However, since Alice cannot link the pseudonyms, it must be cautious of a dishonest vehicle using various pseudonyms to gain an advantage.

Alice needs a method to identify the relationship between such identities.
The most straightforward approach is to send a request to the LEA to confirm that the two received pseudonyms indeed originated from distinct vehicles.
However, maintaining constant contact with LEA is impractical, and such inquiries may compromise privacy.
For these reasons, Alice must avoid making the identification process dependent on LEA involvement.

Instead, if the vehicles include a ``proof of distinctiveness'' in their messages, Alice could verify that received messages are indeed from distinct vehicles.
Moreover, this behavior should not compromise the privacy of sender vehicles.


\subsection{Assumptions}
At the time of writing, the pseudonym is not yet standardized by either ISO or ETSI.
To maintain maximum compatibility with the future, we only make the following two minimum assumptions about the underlying pseudonym system.

Firstly, the pseudonym design should meet the basic common requirements.
Specifically, it should be cryptographically strong enough and have a decent unlinkability.

Secondly, the pseudonym should be centralized issued, i.e. by LEA.
Consequently, the pseudonym will be signed by LEA, with its own public key infrastructure, commonly known as the pseudonym CA.

Lastly, the secret key of pseudonyms should not be extracted or left to the vehicle issued.
This is important because if they can be extracted, then an attacker could gather \textit{permanent} identities from other vehicles with physical access, which is not considered a Sybil attack and thus out of this paper's scope.
This should be ensured by hardware-based security components, such as HSM and anti-tamper OBU.

\subsection{Requirements}
In order to accommodate our work into the existing C-ITS standard framework, some requirements should be met.
A recent survey about detecting and preventing Sybil attacks for C-ITS~\cite{Hammi2022-eb} points out 4 major ``do not'' requirements for a Sybil-resistant system:
\begin{itemize}
    \item Do not modify the existing C-ITS standards.
    \item Do not break current C-ITS security architecture.
    \item Do not rely on vehicles' history.
    \item Do not break the unlinkability of the original pseudonym.
\end{itemize}

Also, it points out several features which are considered adequate for adoption in C-ITS standards:
\begin{itemize}
    \item Do not require additional hardware.
    \item Considers C-ITS standards.
    \item Feasible on current C-ITS systems.
    \item Do not rely on the digital map.
    \item Do not allow tracking by the public.
    \item Scalable to large scenarios.
\end{itemize}

All the above requirements are crucial for the work to be integrated into the standards.
Moreover, we also summarized some additional requirements which are considered equally important:

\begin{itemize}
    \item  Non-interactive: do not require interaction with infrastructure (LEA or RSUs), nor the other vehicles. The former is important because we cannot expect real-time global connectivity in every situation in C-ITS. The latter is also considered good because interacting with other vehicles will increase the traffic and potentially harm the overall decision time.
    \item  Succinct: the work should maintain relatively small communication overheads and low processing footprints for the vehicles.
\end{itemize}

\subsection{Approach}

In accordance with the aforementioned requirements, we propose that vehicles generate cryptographic proof and append it to their messages. The purpose of this proof is to enable Alice to easily verify that two messages received from different pseudonyms were genuinely transmitted by distinct vehicles, rather than being the work of two pseudonyms associated with a single dishonest vehicle. To achieve this, we employ cryptographic techniques. Simultaneously, we prioritize privacy, which motivates the construction of a zero-knowledge proof. The Diophantine equation, known for its challenging solvability, diverse forms, and algebraic properties, serves as a highly suitable zero-knowledge proof quiz in our paper. Consequently, we utilize the Diophantine equation as the foundation of our cryptographic protocol.


The LEA generates a set of equations with identical solutions and attaches each to a pseudonym of the vehicle.
When Charles receives a packet from Alice, it notes Alice's pseudonym $\mathcal{F}$.
Bob, with pseudonym $\mathcal{G}$, is also transmitting a message. 
Alice generates a proof asserting that $\mathcal{F}$ and $\mathcal{G}$ belong to distinct vehicles.
The pseudonym $\mathcal{F}$ is signed by the LEA, and Charles trusts that Alice cannot independently generate the pseudonym $\mathcal{F}$.

Considering another pseudonym $\mathcal{F}'$ associated with Alice, Charles cannot ascertain that it is an additional pseudonym of Alice, as it cannot deduce the solutions of both pseudonyms and identify that they share the same solution.
The security of the aforementioned process is ensured by the implementation of zero-knowledge proofs.

\section{Mathematical Primitives}
\label{sec:math}
Before diving into the actual solution, in this section, we will introduce some mathematical tools to help solve the problem.
The first one is ``the integer solution of a Diophantine equation'', which is considered as the ``computational hard problem'' in the proposed cryptography protocol.
The second one is zk-SNARK, a set of constructions that can be used to convert the first problem into a zero-knowledge proof.

\subsection{Diophantine equation and its integer solution}
\label{subsec:diophantine}
A Diophantine equation is a type of equation involving two or more unknowns with integer coefficients. The general form of a Diophantine equation is given as:
\begin{align}
    \sum_{i=1}^k a_ix_1^{b_{i1}}\dots x_n^{b_{in}}=0.
\end{align}
Where $a_1,\dots,a_n$ are integer, $b_{11},\dots,b_{nn}$ are non-negative integer, and $x_1,\dots,x_n$ are unknowns.

For a given Diophantine equation, let $d = \max{\sum_{j}b_{ij}}$ for all $0<i\leq k$. We refer to such an equation as a $d$-degree Diophantine equation. If $\sum_{j}b_{1j} = \dots = \sum_{j}b_{kj}$ holds, we term it a homogeneous Diophantine equation.

Solving Diophantine equations is trivial when considering solutions in a general sense. However, finding integer solutions to these equations is an exceptionally challenging problem~\cite{andreescu2010introduction}. In the context of this paper, whenever we discuss solutions to Diophantine equations, we are exclusively referring to integer solutions.

The difficulty of solving Diophantine equations increases sharply with the increase in degree. For fourth-degree or higher Diophantine equations, determining whether a solution exists becomes Hilbert's tenth problem, which was solved in 1970 by Matiyasevich. The answer is negative: It is impossible to devise a universal method for such determination~\cite{davis1973hilbert}.

Although finding integer solutions for higher-degree Diophantine equations is extremely hard. Nevertheless, constructing a Diophantine equation with integer solutions is relatively straightforward. Any set of integers can be chosen as the unknowns, and a Diophantine equation with solutions can be generated. Furthermore, verifying whether a set of numbers constitutes a solution to a Diophantine equation is also a trivial task. In other words, the problem of finding solutions to Diophantine equations belongs to the NP-problem. This insight has inspired the use of Diophantine equations in cryptographic designs.

\subsection{zk-SNARK}
In this section, we will introduce a type of zero-knowledge proof known as zk-SNARK~\cite{Ben-Sasson2014-al}.
It enables the prover to show the verifier that he has a solution to an equation without having to reveal the particular solution.

In Section~\ref{sec:related}, we presented an overview of zero-knowledge proofs. In 1991, building upon the foundational work of~\cite{Goldwasser1989-js}, Blum, Feldman, and Micali introduced non-interactive zero-knowledge (NIZK) proofs in their seminal paper~\cite{blum1991noninteractive}. In NIZK proofs, the prover only needs to transmit a single message to the verifier to fulfill the necessary proof requirements.

The following outlines the construction process of zk-SNARKs. Firstly, we articulate the problem that we aim to prove. Consider a polynomial:
\begin{align}
    f(x)=a_1x^p+a_2x^{p-1}\dots a_{p-1}x+a_p.\label{eqn:Poly}
\end{align}
For a specific set of coefficients $a_1\dots a_p$ and a constant $N$, the prover seeks to convince the verifier that they indeed possess a solution $k$ such that $f(k)=N$ without disclosing the exact value of $k$.



\textbf{R1CS \cite{Groth2016-bd}:} Rank-1 Constraint System, or R1CS, provides an alternative representation of the polynomial. Specifically, R1CS rewrites an arithmetic circuit that expresses polynomials into a set of equations. By using R1CS, we can algebraic the arithmetic circuit for further encryption operations.


\textbf{Quadratic Arithmetic Program (QAP) \cite{gennaro2013quadratic}:} QAP is a method for transforming R1CS into a set of vector operations. Initially, we can represent R1CS as a vector $\mathbf{s}$ which constructed by the R1CS.


Assuming there are $n$ wires and $m$ gates in the arithmetic circuit, we construct three $n\times m$ matrices $A,B,C$ such that $A\mathbf{s}\odot B\mathbf{s}=C\mathbf{s}$. Here, $\odot$ denotes the Hadamard product.

Then, we are able to use the Common Reference String (CRS), homomorphic encoding and linear pairing to construct the protocol of zk-SNARK.

The follow-up procedure is described in~\cite{Groth2016-bd}.

In short, zk-SNARK is a cryptographic protocol that allows a prover to convince a verifier that a given statement is true, without revealing any additional information, in a short and efficient manner.
A zk-SNARK protocol consists of three polynomial-time algorithms: a key generator $\mathcal{G}$, a prover $\mathcal{P}$, and a verifier $\mathcal{V}$, defined as follows:

The key generator $\mathcal{G}$ takes as input a security parameter $\lambda$ and a program $C$, and outputs a pair of public keys: a proving key $pk$ and a verification key $vk$. These keys are public parameters that can be reused for multiple proofs for the same program $C$.

The prover $\mathcal{P}$ takes as input the proving key $pk$, a public input $x$ and a private witness $w$. The algorithm outputs a proof $\pi = \mathcal{P}(pk, x, w)$ that attests that the prover knows a witness $w$ such that $C(x,w) = \text{true}$.

The verifier $\mathcal{V}$ takes as input the verification key $vk$, the public input $x$ and the proof $\pi$. The algorithm outputs a binary value $\mathcal{V}(vk, x, \pi) \in \{0,1\}$ that indicates whether the proof is valid or not. The verifier accepts the proof if and only if $\mathcal{V}(vk, x, \pi) = 1$.

\section{Proposed Method}
\label{sec:proposal}

In subsection~\ref{subsec:diophantine}, we have introduced the mathematic open problem, the Diophantine equation.
Based on this, we can construct a cryptographic setup to meet our demands.

Recall that for a high-degree Diophantine equation, finding the solution of it is extremely difficult.
However, if we already know the solution, verifying the solution is easy. 

As an example, Consider a $8$-degree Diophantine equation with 4 unknowns:
\begin{align}
    \mathcal{D}(\vec{a},\vec{x})=a_1x_1^8+a_2x_2^8+a_3x_3^8+a_4x_4^8.
    \label{eqn:D}
\end{align}
Where $\vec{a}=(a_1,a_2,a_3,a_4)$ and $\vec{x}=(x_1,x_2,x_3,x_4)$.
For simplicity of discussion, we decided to use a Diophantine equation with no mixed terms, and the degrees of all unknowns are the same power of $2$.
We chose the $8$-degree Diophantine equation only as an example.
In fact, the degree and number of unknowns can be freely adjusted according to the requirements.
This will be detailed in Section~\ref{sec:discussion}.

By choosing different coefficients $\vec{a}$ we are able to generate different Diophantine equations. For example, we choose an arbitrary set of coefficient $\vec{m}=(m_1,m_2,m_3,m_4)$, then we obtain a Diophantine equation $f(\vec{x})=\mathcal{D}(\vec{m},\vec{x})$.

For an arbitrary set of integers $\vec{s}=(s_1,s_2,s_3,s_4)$, let $y$ be a known constant integer denoting $f(\vec{s})$, we can generate a new Diophantine equation:
\begin{align}
    \mathcal{F}(\vec{s})=m_1s_1^8+m_2s_2^8+m_3s_3^8+m_4s_4^8-y=0.
\end{align}

By doing so, we can obtain an equation with $\vec{s}$ as a solution arbitrarily.
However, it is difficult to solve it to obtain the solution, thus, we can believe that the one who can provide the solution must know $\vec{s}$.

In the same way, if we choose a different set of coefficients $\vec{n}=n_1,n_2,n_3,n_4$, there will be a different Diophantine equation $g(\vec{x})=\mathcal{D}(\vec{n},\vec{x})$. Using the same $\vec{s}$, we obtain $z=g(\vec{s})$ and another Diophantine equation:
\begin{align}
    \mathcal{G}(\vec{s})=n_1s_1^8+n_2s_2^8+n_3s_3^8+n_4s_4^8-z=0.
\end{align}

After the above process, we obtain two Diophantine equations $\mathcal{F}$ and $\mathcal{G}$, which seem unrelated, except that they are both difficult to solve, and only the generator of the equation knows that $\mathcal{F}$ and $\mathcal{G}$ share the same solution $\vec{s}$.

Consider another Diophantine equation $\mathcal{H}(\vec{x})$ which belongs to another vehicle, it is easy to verify that $\mathcal{H}(\vec{s})\neq 0$ since that $\mathcal{H}$ is generated by a set of integers which different from $\vec{s}$ which is guaranteed and assigned by LEA.

By using the above method, we can generate any number of Diophantine equations with known solutions.
The solution $\vec{s}$ is the fixed and unique identity of the creator or prover.
For the verifier, we need to construct a method such that they can believe that the prover has the solution of a Diophantine equation, and the solution is not another equation's solution.
In other words, they need to verify that the prover has a vector $\vec{s}$ that satisfies the following set of equations:
\begin{equation}
    \begin{cases}
    \mathcal{F}(\vec{s})=0\\
    \mathcal{G}(\vec{s})\neq0
    \end{cases}
\end{equation}

We refer to the above process as the \textbf{distinct identity criterion (DIC)}.

Notice that, since the pseudonyms have enough entropy, we can let every party generate the $\vec{a}$ via a determined algorithm (e.g. cryptographic hash) on the fly, to get rid of transferring them along with the pseudonyms.
The only variable that needs to be transferred (and sealed) is the constant term $y$.

\hfill\\

Next, to enable a prover to show that it possesses the solution $\vec{s}$ without directly revealing the solution themselves, the equations in ``plaintext'' should be converted to an arithmetic circuit, which can in turn be converted to prove function (PF) and verify function (VF) of zk-SNARK.

\begin{figure}[ht]
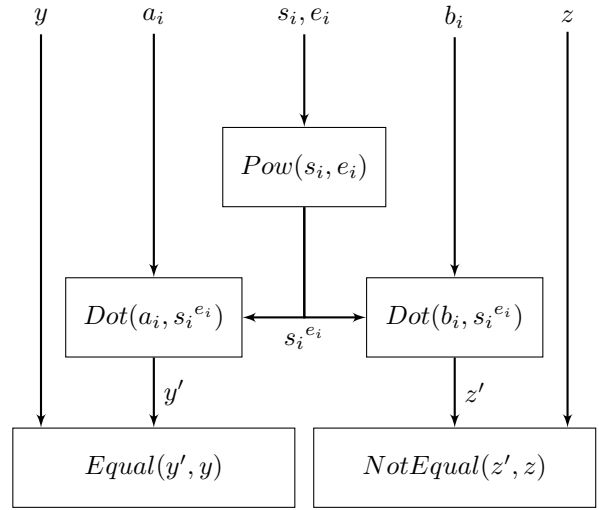

    \centering
    \include{figures/circuit}
    \caption{Arithmetic circuit overview}
    \label{fig:circuit}
\end{figure}

The overview of the corresponding arithmetic circuit is depicted in Figure~\ref{fig:circuit}.

Function $Pow(s, e)$ is to calculate the $s^e$ with integer $s$ and $e$ using the Montgomery algorithm~\cite{Montgomery1985-di}. 
In the situation of the degree of Diophantine equations being a power of $2$, the total number of multiplication gates will be $log_2 e$.

Function $Dot(A, B)$ is the trivial dot product circuit, which calculates the pairwise product of each item in two vectors and sums all the products together.
The number of multiplication gates is $|A|$.

Function $Equal(a, b)$ is directly translated to a multiplication gate in R1CS, namely $1 \cdot a = b$.

Function $NotEqual(a, b)$ requires a slightly complicated construction and can be translated to a multiplication gate with the form of $v \cdot (a - b) = 1$.
This is possible because: if $a \neq b$, there must be a free variable $v$ which satisfies the constraint; if $a = b$, then $a - b = 0$, and nothing can make $v \cdot 0 = 1$.


\hfill \\

Utilizing the DIC, protected by zk-SNARK, we finally propose the actual cryptographic protocol, called \textbf{zero-knowledge Proof of Distinct Identity (zk-PoDI)}, which is outlined as follows:

\noindent\textbf{Setup, done by LEA:} 

\begin{enumerate}

\item Global setup: \\
    Pick a complex enough form of Diophantine equations $\mathcal{D}$.
    Express the corresponding $DIC$ in the language of R1CS and calculate the corresponding arithmetic circuit $C$.
    Generate the proof function $PF$ and verify function $VF$ based on $C$ with the corresponding keys $(pk,vk)$, based on zk-SNARK.
    Distribute $(pk,vk)$ to every vehicle.

\item Generate the orthonym for each vehicle: \\
    Pick a set of cryptographic random numbers $\vec{s}=(s_1,\dots,s_n)$ as the prover's orthonym.
    Distribute each orthonym to the vehicles.

\item Generate each pseudonym of each vehicle: 
    \begin{enumerate}

    \item Generate a sufficient list of pseudonyms $P_1,\dots$ for the vehicle, using the underlying pseudonym generation algorithm.

    \item Generate coefficients $\vec{m}_k=(m_{k1},\dots,m_{kn})$ based on the pseudonym $P_k$ of the vehicles. The particular generation algorithm is not limited as long as it is strong enough, e.g. using cryptographic hash functions.

    \item For a pseudonym $P_k$, plug in the $\vec{s}$ and $\vec{a}$ into quiz generation method to generate the quiz equation $\mathcal{F}(x)=0$. It is determined by $P_k$ and $y_k=f(\vec{s})$ so everyone can reproduce $\mathcal{F}$ according to them.

    \item Sign each pair of $P_k$ and $y_k$ with the pseudonym CA and distribute them to the vehicle.
    \end{enumerate}
\end{enumerate}

\noindent\textbf{Prove:}

\begin{enumerate}
    \item Given the current pseudonym $P_e$, $y_e$ of the egovehicle, and the signature-verified pseudonym $P_k$, $y_k$ of another vehicle.
    \item Generate coefficients $\vec{a}_e$ according to the pseudonym $P_e$; generate $\mathcal{F}$ using $\vec{a}_e$ and $y_e$.
    \item Do the same with $\vec{a}_k$ and $y_k$ to obtain $\mathcal{G}$.
    \item Calculate $\pi=PF(pk;\mathcal{F},\mathcal{G};\vec{s})$.
    \item Publish $\pi,P_e,y_e,P_k,y_k$.
\end{enumerate}

\noindent\textbf{Verify:}

\begin{enumerate}
    \item Given $\pi,P_e,y_e,P_k,y_k$.
    \item Verify the LEA signature of $P_e,y_e$ and $P_k,y_k$.
    \item Generate the quiz equations $\mathcal{F}$ and $\mathcal{G}$ according to the pseudonym $P_e,P_k$.
    \item Accept if $VF(vk;\mathcal{F},\mathcal{G};\pi)=true$, otherwise reject.
\end{enumerate}

\begin{figure}[ht]
    \centering
    \includegraphics[width=.9\linewidth]{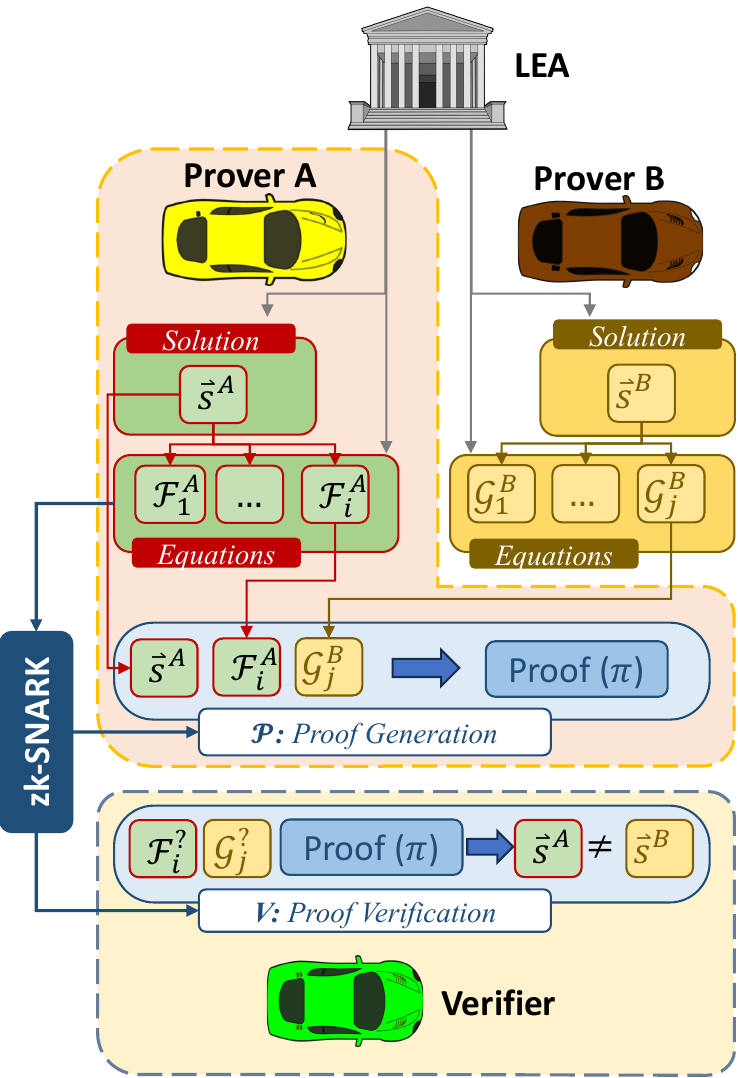}
    \caption{Protocol overview}
    \label{fig:protocol}
\end{figure}

Figure~\ref{fig:protocol} depicts an overview of how the protocol works and what is transmitted from LEA to the vehicles, from prover to verifiers.
The $pk$ and $vk$ are omitted in the figure.
LEA creates the solution for each prover as its orthonym $\vec{s}$; then it creates one quiz equation $\mathcal{F}_1, \mathcal{F}_2, \cdots$ for each of its pseudonyms, and distributes all of them to the vehicle with LEA's signature (endorsement.)
For a prover, it plugs in its own $\vec{s}$, the quiz function of its current pseudonym, $\mathcal{F}_i$, and the function of the other vehicle, $\mathcal{G}_j$. The output is a value $\pi$, representing its proof of distinctiveness.
Finally, for a verifier, it can be convinced that $\mathcal{F}_i$ and $\mathcal{G}_j$ are from distinct vehicles by using only $\mathcal{F}_i$, $\mathcal{G}_j$ and $\pi$.

\hfill \\

The zk-PoDI is considered a valid zero-knowledge proof system because it satisfies all three required properties:

\noindent\textbf{Completeness:}
If a dishonest prover conducts a Sybil attack, i.e. they use two different pseudonyms and pretend to be two different vehicles. In this case, they cannot construct a solution $\vec{s}$ such that the DIC is held, because the two equations are both true (the real solution) or both false (the non-solution). 

\noindent\textbf{Soundness:}
Solving the Diophantine equation is an extremely difficult problem. With today's computational techniques, it is still impossible to solve a high-degree Diophantine equation. With the increasing of the amount of unknown numbers, the probability of a random collision solution drops to close to zero.

The constant term $y$ is signed by the LEA together with the pseudonym $P$.
Although the coefficients $\vec{m}$ are generated on the fly, however, $\vec{m}$ is generated by the $P$ which is still signed by the LEA.
Therefore, a dishonest vehicle cannot construct a separate equation since the whole equation is sealed by LEA.

Finally, the existence of such $\vec{s}$, which indicates the prover indeed possesses the orthonym, is guaranteed by zk-SNARK.

\noindent\textbf{Zero-Knowledge:}
The protocol is zero-knowledge because the orthonym $\vec{s}$ is the secret input of zk-SNARK and thus will not be revealed.
Instead, the proof $\pi$ carries all the information needed to convince a verifier.

\section{Discussion}
\label{sec:discussion}

\subsection{Performance analysis}

Our protocol uses the Groth16~\cite{Groth2016-bd} implementation of zk-SNARKs because of its minimum verifier and communication complexity.
Groth16's major downside is it requires a per-circuit setup procedure, which might be a problem in other applications.
However, this is not a problem for our zk-PoDI, because it only requires one global circuit for DIC during the deployment of the whole pseudonym system.

Based on the previously mentioned assumptions, we can fix the form of DIC as Equation~\ref{eqn:dic-experiment} and analyze the complexity of zk-PoDI with different numbers of equations ($Nx$) and numbers of power ($Np$).
Here we limit the $Np$ to be a power of $2$, to achieve maximum efficiency in the Montgomery algorithm.
By adjusting the super-parameters $Nx$ and $Np$, we can evaluate how the strength of zk-PoDI against different attacks changes.

\begin{equation}
  \begin{cases}
    \mathcal{F}(\vec{s})=a_1s_1^{Np}+a_2s_2^{Np}+\cdots+a_{Nx}s_{Nx}^{Np}-y=0 \\
    \mathcal{G}(\vec{s})=b_1s_1^{Np}+b_2s_2^{Np}+\cdots+b_{Nx}s_{Nx}^{Np}-z\ne0 
  \end{cases}
  \label{eqn:dic-experiment}
\end{equation}

In our implementation, we use the Alt-bn128 curve, which is a commonly used curve in e.g. the Ethereum blockchain.
The finite fields $\mathbb{F}_q$ and $\mathbb{F}_{q^2}$ in Alt-bn128 is $254$ and $508$ bits long;
the element of $\mathbb{G}_1$ is $\mathbb{F}_q$ plus a sign bit;
the element of $\mathbb{G}_2$ is $\mathbb{F}_{q^2}$ plus a sign bit;
and the proof consists of two elements in $\mathbb{G}_1$ and one in $\mathbb{G}_2$.
Thus, the total size of the proof $\pi$ is given by:
$$|\pi| = 2|\mathbb{G}_1| + |\mathbb{G}_2| = 2 (254 + 1) + 508 + 1 = 1019 \mathrm{bits}$$

We implemented the generator, prover, and verifier functions of zk-PoDI using the de-facto zk-SNARK library \textit{libsnark}, and conducted experiments in an Ubuntu 22.04 workstation with Intel i9-13900KF CPU and 128GB DDR4 2133 DRAM.
For each experiment, we repeated the prover and verifier functions 1000 times each and calculated the average proof and verification time.

\begin{figure}
    \centering
    \includegraphics[trim={0 0 0 3.5cm},clip,width=\linewidth]{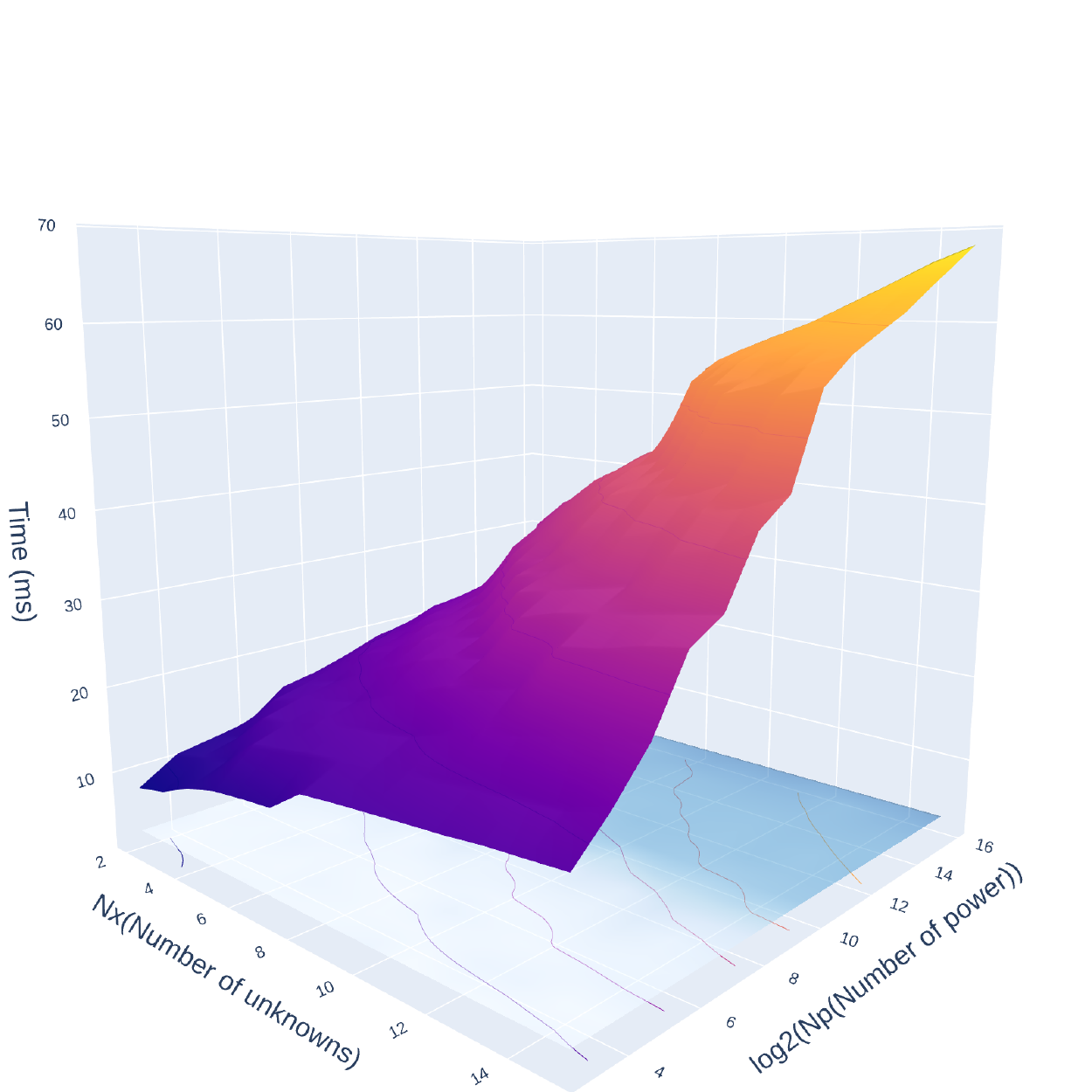}
    \caption{Proof and verification time of zk-PoDI}
    \label{fig:podi-proof-time}
\end{figure}

The average proof time $T_p$ and verification time $T_v$ against $Nx$ and $Np$ are plotted in Figure~\ref{fig:podi-proof-time}.
The $T_p$ is shown in the upper plane with purple to yellow colors, and its contour is projected into x-y plane.
The $T_v$ is shown in bluish color at the bottom of the figure for comparison.
From the plot, we can see that the $T_v$ is nearly constant at approximately $3$ ms, but the $T_p$ varies from $8$ to $72$ ms.
The $T_p$ has a positive relationship with $Nx$ and $Np$ respectively, and increasing $Nx$ has a similar effect of increasing $log_2(Np)$.
This feature gives zk-PoDI the ability to freely adjust the ratio of $T_p/T_v$, which is considered convenient in designing the actual applications.
Also, with high enough $T_p/T_v$, zk-PoDI can be treated as a protocol with an embedded proof of work (PoW), which can then mitigate various attacks including denial of service (DoS) or brute force attacks, conceptually similar to the key derivation function (KDF).

While in practice, the number of surrounding vehicles is usually greater than one.
A possible optimization could be done by proving the distinctiveness of several adjacent vehicles at a time.
The corresponding DIC is as follows and the modification to the circuit is obvious:
\begin{equation}
    0=\mathcal{F}(\vec{s})\neq\mathcal{G}(\vec{s})\neq\mathcal{H}(\vec{s})\neq\cdots
\end{equation}

\subsection{Threat analysis}
From the ZKP properties of the protocol, we directly know that a malicious prover cannot make the verifier convinced.
However, it still may generate many false proofs to exhaust the verifier's computation power, which is known as DoS attack.
Nevertheless, from the performance analysis, the $T_p/T_v$ can be set to a high enough value to make DoS attacks considered non-profitable.

Potential attacks based on the malleability of equations, including chosen plaintext attack, chosen identity attack, etc. are not possible because the coefficients of equations are directly generated from the pseudonym and the constant term is signed in conjunction with the pseudonym by the LEA.


Another potential attack is, since the pseudonyms (thus also equations) may be preloaded into the vehicles, an attacker may attempt to find solutions other than $S$ in an offline fashion, with GPUs or ASICS.
A possible mitigation is to quantize the difficulty of finding solutions and select an equation to make the expectation solving time much longer than the lifetime of a pseudonym.

\subsection{Difficulty analysis}

As discussed earlier, solving a general Diophantine equation is known to be a problem that cannot be efficiently addressed in polynomial time~\cite{Hilbert10}, ensuring the mathematical security of our protocol. Despite this, numerous new methods have been proposed to attempt solving Diophantine equations within acceptable time frames. In recent years, optimization methods have become prevalent in addressing Diophantine equations, with some proving effective, such as ant colony optimization~\cite{abraham2013finding} and connectionist-based models~\cite{jeswal2020connectionist}.

However, while these methods demonstrate their effectiveness in solving specific Diophantine equations, they are still far from achieving polynomial efficiency. Our protocol has the capability to generate a substantial number of high-order multi-coefficient Diophantine equations. Consequently, even though these methods exhibit rapid solving speeds for simpler Diophantine equations, they do not pose a threat to the security of our protocol.

Nonetheless, it remains crucial to monitor the progress of the solutions that are continually being developed. If a method emerges that can efficiently solve a particular class of Diophantine equations in polynomial time, it is advisable to exclude such easily solved equations from our equation generation process. Since our choice of equations is arbitrary, taking such a precautionary step is acceptable.

\section{Conclusion}
\label{sec:conclusion}

In this work, we proposed the cryptographic protocol zero-knowledge Proof of Distinct Identity (zk-PoDI,) which enables the vehicles to prove that it is not the owner of another known pseudonym in the local area.
Such a mechanism solved the paradox: it prevents the Sybil attack while maintaining the unlinkability of the pseudonym.

According to a recent survey~\cite{Hammi2022-eb}, zk-PoDI fully fulfills all the requirements regarding a practical Sybil-resistance pseudonym system.
zk-PoDI features zero latency, moderate computation overhead, and negligible communication cost.
More importantly, it is designed to work independently and statelessly.
It does not rely on a particular design of the underlying pseudonym system, does not rely on extra knowledge about the environment, the historical behavior of other vehicles, or any sort of assistance from the infrastructure.
These features make the zk-PoDI very distinctive compared to the existing works and we believe that it is not far from being integrated into the C-ITS system.

However, we admit that the evaluation in a realistic environment is necessary, to verify the protocol itself, and its overall impact on the system.
Therefore, in the future, we will implement the zk-PoDI in the Flowsim simulation platform~\cite{Tao2023-oe}, with multiple pseudonym implementations and integration with different C-ITS applications, and evaluate it in a city-scale realistic environment.

\section*{Acknowledgment}
These research results were obtained from the commissioned research Grant number 01101 by the National Institute of Information and Communications Technology (NICT), Japan.

\bibliographystyle{unsrt}
\bibliography{tydus,xana}

\end{document}

%% file: figures/circuit.tex
\begin{tikzpicture}[auto, node distance=2cm,>=latex']

\node [text width=5em, text centered, minimum height=1em,name=genesis] (S) {$s_i,e_i$};

\node [text width=5em, text centered, minimum height=1em, left of=S, node distance=2cm] (A) {$a_i$};
\node [text width=5em, text centered, minimum height=1em, left of=A, node distance=1.5cm] (Y) {$y$};

\node [text width=5em, text centered, minimum height=1em, right of=S, node distance=2cm] (B) {$b_i$};
\node [text width=5em, text centered, minimum height=1em, right of=B, node distance=1.5cm] (Z) {$z$};

\node [rectangle, draw, text width=5.5em, text centered, minimum height=3em, below of=S,node distance=2cm] (Pow) {$Pow(s_i,e_i)$ };
\draw [->,thick] (S) -- node {} (Pow);

\node [rectangle, draw, text width=6em, text centered, minimum height=3em, below left of=Pow,node distance=2.82842712474619cm] (Dot1) {$Dot(a_i, {s_i}^{e_i})$};
\node [rectangle, draw, text width=6em, text centered, minimum height=3em, below right of=Pow,node distance=2.82842712474619cm] (Dot2) {$Dot(b_i, {s_i}^{e_i})$};

\draw [->,thick] (A) -- node {} (Dot1);
\draw [->,thick] (Pow) |- node {${s_i}^{e_i}$} (Dot1);
\draw [->,thick] (B) -- node {} (Dot2);
\draw [->,thick] (Pow) |- node {} (Dot2);

\node [rectangle, draw, text width=10em, text centered, minimum height=3em, below of=Dot1,node distance=2cm] (Eq) {$Equal(y', y)$};
\node [rectangle, draw, text width=10em, text centered, minimum height=3em, below of=Dot2,node distance=2cm] (NEq) {$NotEqual(z', z)$};

\draw [->,thick] (Y) -- node {} ([xshift=-1.5cm]Eq.north);
\draw [->,thick] (Dot1) -- node {$y'$} (Eq);
\draw [->,thick] (Z) -- node {} ([xshift=1.5cm]NEq.north);
\draw [->,thick] (Dot2) -- node {$z'$} (NEq);

\end{tikzpicture}